\documentclass[lettersize,journal]{IEEEtran}

\usepackage{cite}
\usepackage{amsmath,amssymb,amsfonts}

\usepackage{graphicx}
\usepackage{textcomp}
\usepackage[utf8]{inputenc} 
\usepackage[T1]{fontenc}    
\usepackage{hyperref}       
\usepackage{url}            
\usepackage{booktabs}       
\usepackage{amsfonts}       
\usepackage{nicefrac}       
\usepackage{microtype}      
\usepackage{graphicx}
\usepackage{tabularx}
\usepackage{hyperref}
\usepackage{color}
\usepackage{amsmath}

\usepackage{algorithm}  
\usepackage[noend]{algpseudocode} 
\usepackage{subfigure}
\usepackage{multirow}
\usepackage{soul}
\begin{document}

\title{The Balanced Matrix Factorization for Computational Drug Repositioning}
\author{Xinxing~Yang, \IEEEmembership{} Genke~Yang, and Jian~Chu \IEEEmembership{}
	\thanks{This research was partially funded by the China National R\&D Key Research Program (2020YFB1711204) (2019YFB1705702). (Corresponding Author: Genke~Yang; Email: gkyang@sjtu.edu.cn.)}
	\thanks{X. Yang, G.Yang and J.Chu are with the the Department of Automation, Shanghai Jiaotong University, Shanghai 200240, China (e-mail: yangxinxing@sjtu.edu.cn; gkyang@sjtu.edu.cn; chujian@sjtu.edu.cn).}}

\maketitle

\begin{abstract}
	Computational drug repositioning aims to discover new uses of drugs that have been marketed. However, the existing models suffer from the following limitations. Firstly, in the real world, only a minority of diseases have definite treatment drugs. This leads to an imbalance in the proportion of validated drug-disease associations (positive samples) and unvalidated drug-disease associations (negative samples), which disrupts the optimization gradient of the model. Secondly, the existing drug representation does not take into account the behavioral information of the drug, resulting in its inability to comprehensively model the latent feature of the drug. In this work, we propose a balanced matrix factorization with embedded behavior information (BMF) for computational drug repositioning to address the above-mentioned shortcomings. Specifically, in the BMF model, we propose a novel balanced contrastive loss (BCL) to optimize the category imbalance problem in computational drug repositioning. The BCL optimizes the parameters in the model by maximizing the similarity between the target drug and positive disease, and minimizing the similarity between the target drug and negative disease below the margin. In addition, we designed a method to enhance drug representation using its behavioral information. The comprehensive experiments on three computational drug repositioning datasets validate the effectiveness of the above improvement points. And the superiority of BMF model is demonstrated by experimental comparison with seven benchmark models.
\end{abstract}


\footnote{Preprint. Work in progress.}

\section{Introduction}
\label{sec:introduction}
\IEEEPARstart{D}{rug} development is a long, costly and failure-prone process \cite{1}. In traditional small molecule drug development, for example, it takes three stages to design a successfully marketed drug for a disease from scratch: drug discovery, preclinical studies, and clinical studies \cite{2,3}. Each of these stages is subdivided into many smaller stages, such as drug discovery consisting of target identification, design of lead compounds, and other processes. Each of the above steps is fraught with uncertainty and consumes a great deal of time and money \cite{4,5}. Furthermore, current drug discovery strategies are unable to cope with sudden outbreaks, such as the current COVID-19 virus. Therefore, we need a new drug discovery technology that can accelerate drug development and reduce the cost of R\&D \cite{6}. \par

The purpose of computational drug repositioning is to discover new uses of drugs that have been marketed or are in the clinical stage beyond their original indications \cite{7}. In contrast to traditional drug development strategies that require designing a new drug from scratch, computational drug repositioning approaches only require a priori knowledge of a large number of known drug-disease associations, the molecular structure of the drug, and medical information about the disease, and then can be used to infer new uses for known drugs through data mining algorithms or artificial intelligence algorithms \cite{8}. The advantage of this technology is that it can save time and money by eliminating some of the traditional drug development processes. And because the marketed drugs have passed rigorous clinical trials and obtained the approval of the US Food and Drug Administration, the possibility of their subsequent listing is high, which reduces the risk \cite{9,10}. \par

Current computational drug repositioning methods include matrix factorization-based models and graph-based models \cite{11,12,13,14,15,16}. The graph-based model begins with the construction of a drug-disease association network. This is followed by information mining of this network using graph algorithms to infer potential drug-disease associations \cite{n1,n2,n3}. Luo et al. \cite{rw11} proposed a new computational drug repositioning model based on the random walk algorithm. The model first constructs a heterogeneous network containing drugs and diseases, where the edge between drug nodes represents the similarity between these two drugs, the edge between disease nodes represents the similarity between these two diseases, and the presence of edges between drug nodes and disease nodes represents the existence of a therapeutic relationship between them, and the absence of edges represents the absence of a therapeutic relationship. The potential drug-disease association is then mined by the Bi-Random walk algorithm. Chen et al. \cite{rw12} borrowed cross-domain recommendation and transfer learning in recommender systems to solve the problem of computational drug repositioning and drug-target association prediction at the same time. The benefit of this method is that by simultaneously considering two association networks, the drug-disease association network and the drug-target association network, a more informative drug representation can be learned, which can enhance its expressive power in the prediction stage. A model \cite{rw13} based on graph convolutional networks was proposed by Cai et al. The model first constructs a heterogeneous network using molecular information about the drug, medical information about the disease, and known drug-disease associations. They then used a graph convolutional neural network and a novel feature extraction module to mine the topological information of drugs and diseases on this heterogeneous network to be able to obtain the representation of the drug or disease. And to enhance the expressiveness of the representations of drugs or diseases, they employ an attention mechanism to integrate these representations. Subsequent experimental results verify the effectiveness of the model. Ge et al. \cite{rw14} designed a computational drug repositioning model for the discovery of potential drugs capable of treating COVID-19. 6255 drugs were considered for repositioning including marketed drugs, drugs in development, and experimental drugs. The method first uses public data to construct a knowledge graph containing node types such as drugs, viral targets, and human targets. The original 6,255 drugs were then narrowed down to 212 using a graph-based mining algorithm. Then, through literature filtering and other methods, drugs that may have a therapeutic relationship with the COVID-19 are finally inferred. \par

The matrix factorization-based model \cite{m1,m2,m3} is divided into two main steps, the first step is to map the drug or disease to the corresponding embedding space by coding technique and represent the drug or disease with a dense numerical vector, which is called the latent factor of the drug or disease. So the matrix factorization-based model can also be called the latent factor-based model. The compatibility between the latent factor of the drug and the latent factor of the disease is then calculated using the similarity function. The most classic matrix factorization model \cite{rw21} is to use two low-rank matrices to represent the latent factors of drugs and diseases respectively and use the inner product operation as the similarity function. These two low-rank matrices are trained with the loss between the labeled data and the predicted values, thus allowing them to make predictions for new drug-disease associations. Chen et al. \cite{rw22} argue that the existing methods suffer from data sparsity, causing a decrease in their prediction performance. And the existing fusion strategies cannot fuse the information effectively. Therefore, they propose a new computational drug repositioning model. The model utilizes deep autoencoders to perform feature extraction on structural data of drugs and diseases to obtain better latent factors. Then, the similarity information of drugs and diseases is calculated from various data. Finally, a variety of data are fused through an adaptive fusion strategy to obtain the predicted drug-disease association probability. For the matrix factorization model does not obey the triangle inequality, making it unable to effectively capture the finer-grained preferences of drugs and diseases, Luo et al. \cite{rw23} propose a new method for computational drug repositioning based on collaborative metric learning that can more intuitively characterize the association between drugs and diseases. The method uses metric learning to mine the respective latent factors of drugs and diseases, and then uses these latent factors to calculate the probability of a therapeutic association between drugs and diseases. Similar to Luo et al.'s work, Yang et al. \cite{m3} also proposed a neural metric factorization for predicting potential drug-disease associations by addressing the problem that matrix factorization models do not obey triangular inequalities. Different from Luo et al.'s method of using metric learning, they regard drugs and diseases as a point in the embedding space, and use an improved Euclidean distance to represent the relationship between the drug and disease. In their article, they showed how expressions such as point and Euclidean distance fit into triangle inequalities. \par

However, the existing research work is plagued by the following problems. Firstly, in the real world, only a small number of diseases have definite treatment drugs, and most diseases have no definite treatment drugs. Therefore, the ratio of validated drug-disease associations (positive samples) and unvalidated drug-disease associations (negative samples) in the dataset of computational drug repositioning is imbalanced, and the ratio is about 1:100. The imbalance in the number of positive and negative samples causes the model to learn a priori information about the imbalance ratio, which affects the model's ability to learn the more essential association laws \cite{focaloss}. It makes the model tend to predict the unknown drug-disease association samples as negative samples when making predictions, reducing its robustness. The reason for this phenomenon is that due to the imbalance of positive and negative samples, when the model trains the parameters it contains, the gradient of the negative samples will dominate the optimization direction of the loss function. Moreover, the negative sampling technique samples a large number of negative samples, which will have redundant samples (useless information), and the optimization gradient of the model will be overwhelmed by these uninformative samples, thereby reducing the performance of the model. Lastly, due to the imbalance in the number of positive and negative samples, the number of high-confidence samples is relatively large, which ultimately dominates the gradient in the loss function. However, high-confidence samples have very little improvement on the model performance, and the model should focus on improving the gradient of low-confidence samples in the loss function. \par

Secondly, the existing work uses information such as the small molecule structure and chemical formula of the drug as the features of the drug, and does not consider the behavioral information of the drug, which makes it inability to comprehensively model the intrinsic characteristics of the drug. In the field of recommender systems, many works have demonstrated that using user behavior information can enhance its representation ability, thereby improving the generalization ability of the model. Therefore, how to model the behavior of drugs is the key to improving the performance of models. Going with the flow, we have been inspired by studying solutions to these problems in the fields of computer vision and recommender systems \cite{focaloss,cl,simplex}.\par

\begin{figure*}[t]
	\centering
	\includegraphics[scale=0.7]{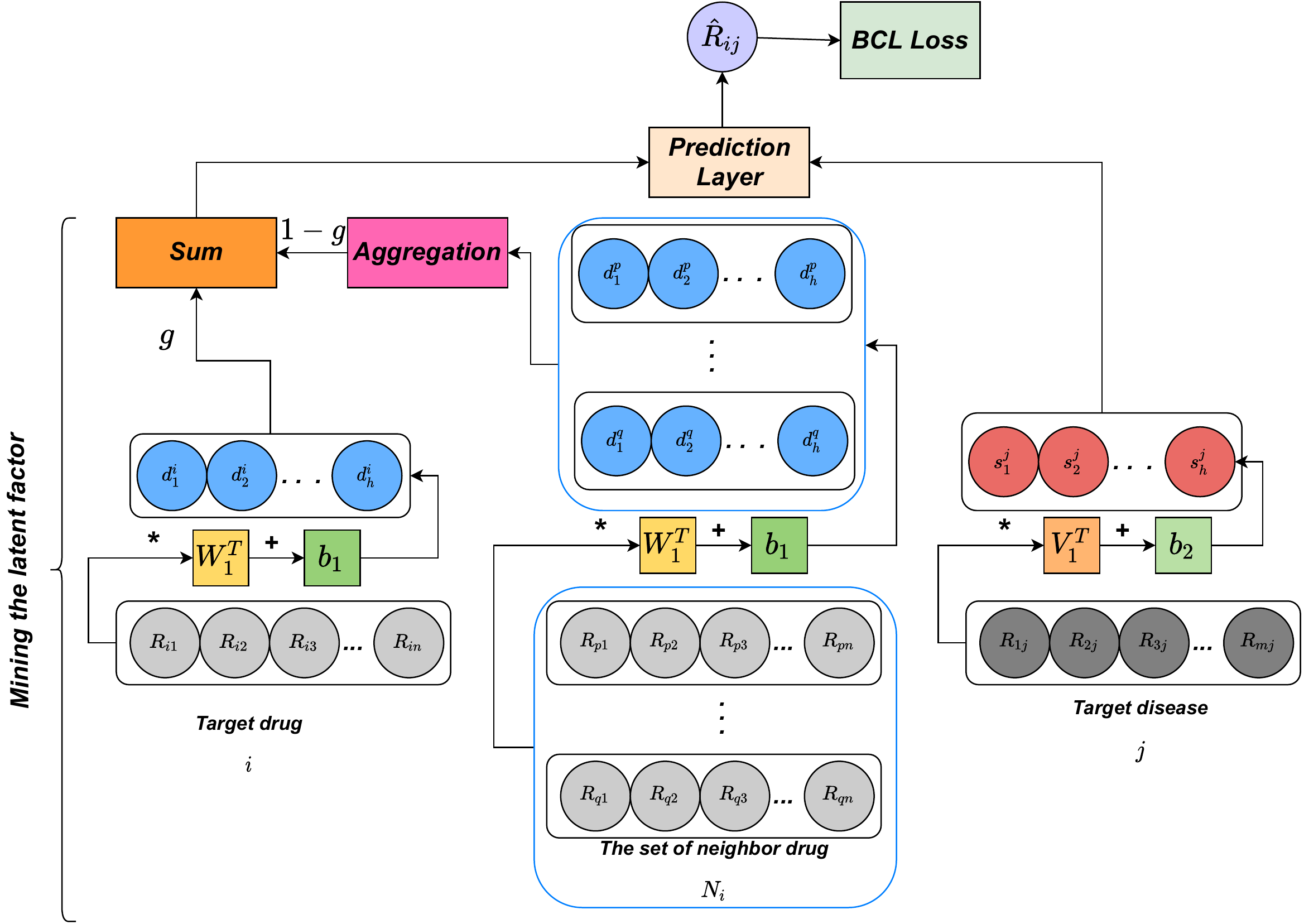}
	\caption{The architecture of the BMF model.}
	\label{}
\end{figure*} 

In this work, we propose a balanced matrix factorization with embedded behavior information (BMF) for computational drug repositioning to address the above-mentioned shortcomings. Specifically, in the BMF model, we propose a novel balanced contrastive loss (BCL) to optimize the category imbalance problem in computational drug repositioning. The BCL optimizes the parameters in the model by maximizing the similarity between the drug's latent factor and disease's latent factor (with a therapeutic relationship), and minimizing the similarity between the drug's latent factor and disease's latent factor below the margin (without a therapeutic relationship). The novelty of this loss function is that firstly BCL filters uninformative negative samples by setting the minimum margin so that their gradient in the loss is zero, which can accelerate the convergence of the model without being disturbed by these redundant samples. Secondly, BCL improves the generalization ability of the model by reducing the weight of high-confidence samples in the loss function, which can make the model focus on improving the accuracy of samples with low confidence. Thirdly, BCL solves the gradient problem arising from the imbalance between positive and negative samples by adjusting the weights of positive and negative samples in the loss function through a controllable parameter. \par

In addition, in the BMF model, we designed a method to enhance drug representation using behavioral information. Specifically, when making association predictions for target drug $i$ and target disease $j$, the representation of target drug $i$ considers behavioral information from both the drug and the disease. Firstly, the input used to mining the latent factor drug $i$ is not one hot vector, but the treatment record vector of the drug for all diseases (drug behavior information). And all drugs that are therapeutically associated with the target disease $j$ are regarded as the neighbors of the target drug $i$, and then the representation of the target drug $i$ and the representation of the neighbor drugs are merged through a fusion strategy (disease behavior information). In this way, multiple behavioral information can be embedded into the representation of the target drug $i$. The ablation experiments on three real datasets verify the effectiveness of the above strategies. And the BMF model outperforms the current SOTA model under three evaluation metrics.\par

The main contributions made by this work are as follows: \par

\begin{enumerate}
	
	\item We propose a balanced matrix factorization with embedded behavior information (BMF) for enhanced drug representation and applies to computational drug repositioning scenarios where positive and negative samples are imbalanced.
	
	\item We propose a new balanced contrastive loss function, BCL, for optimizing category imbalance problem in computational drug repositioning. Specifically,  BCL solves the accuracy degradation caused by the imbalance of positive and negative samples by setting the minimum margin to filter uninformative negative samples, reducing the loss weights of high confidence samples and adjusting the gradient weights of positive and negative samples.
	\item We design a novel approach to augment drug representations with behavioral information. Specifically, in the process of mining drug representations, the disease information associated with the drug and the representation information of neighboring drugs are additionally considered, to obtain a better drug representation.
	\item Through ablation experiments on three real computational drug repositioning datasets, we validated the effectiveness of BCL and behavioral information. We also demonstrate the superior performance of the BMF model by comparing six popular computational drug repositioning models.
\end{enumerate}

\begin{figure*}[t]
	\centering
	\includegraphics[scale=0.7]{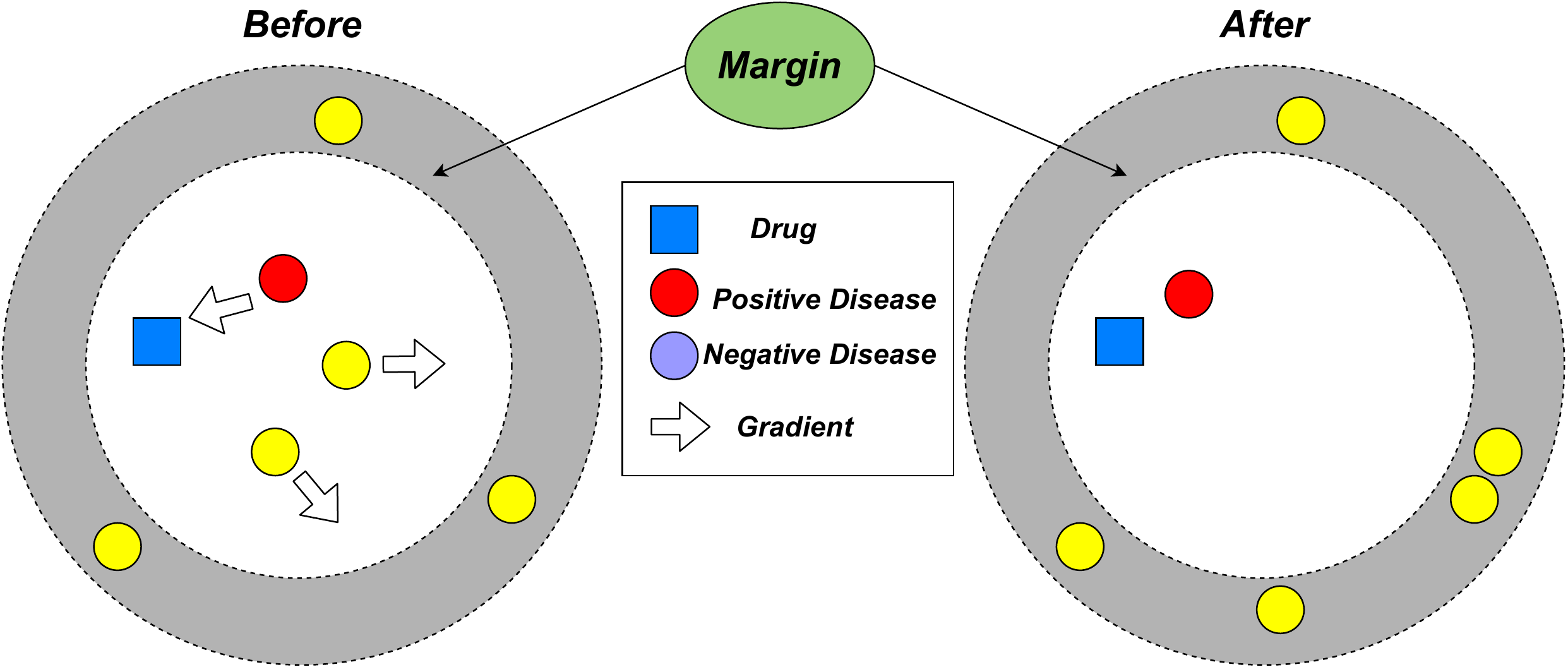}
	\caption{The comparison of the latent factor (drugs and diseases) after BCL optimization.}
	\label{}
\end{figure*} 

The remaining content of this work is organized as follows. The dataset used in this work will be introduced in section \uppercase\expandafter{\romannumeral2} and the details of the BMF model will be introduced in section \uppercase\expandafter{\romannumeral3}. In the section \uppercase\expandafter{\romannumeral4}, the experimental setup, ablation experiments and comparison experiments will be introduced. The section \uppercase\expandafter{\romannumeral5} discusses the strengths and weaknesses of this work, and discusses future work directions. \par

\begin{table}[h]
	\caption{The statistics of datasets used in this study}
	\centering
	\setlength{\tabcolsep}{2.5mm}
	\begin{tabular}{@{}ccccc@{}}
		\toprule
		Datasets  & Drugs & Diseases & Validated Associations & Sparsity \\ \midrule
		Gottlieb  & 593   & 313      & 1933                   & 98.95\%   \\
		Cdataset  & 663   & 409      & 2532                   & 99.06\%   \\
		DNdataset & 1490  & 4516     & 1008                   & 99.98\%   \\ \bottomrule
	\end{tabular}
\end{table}

\section{Datasets}
We first describe the statistics of the datasets used in this work. We used three publicly available benchmark datasets, Fdataset \cite{11}, Cdataset \cite{13} and DNdataset \cite{rw11}. These datasets have differences in the number of drugs, the number of diseases, and the number of validated associations, which can comprehensively verify the true performance of the model. As shown in table \uppercase\expandafter{\romannumeral1}, the number of drugs in the Fdataset, Cdataset, and DNdataset dataset is 593, 663, and 1490, the number of diseases is 313, 493, and 4516, and the number of known validation associations is 1933, 2532, and 1008, respectively. \par
All the drugs in the above dataset are approved by the FDA, and their specific information, including their chemical structural formula, can be queried in DrugBank \cite{drugbank}. All diseases in the above dataset can be queried from the OMIM database \cite{omim}. It is worth noting that the method proposed in this paper does not require drug and disease similarity information \cite{cdk,smile}. Since there are a large number of drugs and drugs with no known therapeutic relationship in the DNdataset, we removed the drugs without any treatment of the disease, and the diseases without any drug that can be cured, and finally 550 drugs and 360 diseases were retained in the DNdataset.\par

\section{Methods}

In this section, the BMF model proposed in this work will be described, and the balanced contrastive loss and embedded behavioral information will be discussed. \par

\subsection{The Balanced Matrix Factorization with Embedded Behavior Information}

In this study, we propose the balanced matrix factorization with embedded behavior information (BMF) for computational drug repositioning. The matrix factorization-based method uses one-hot coding to model the latent factor of drug and disease, and then uses the similarity function to calculate the probability of the association between drug and disease. In the training phase, the model uses the error between the predicted data and the labeled data to train the latent factor of drug and disease. In the prediction stage, the trained latent factor of drug and disease can predict whether there is an association between potential drug and disease. \par

Figure 1 shows the framework of the BMF model which is mainly extended on the idea of matrix factorization. From bottom to top, it shows the process of mining latent factor of target drug $i$ and target diseases $j$, and the process of calculating predicted values $\hat{R}_{ij}$. The innovation of the BMF model is that it embeds behavioral information in the process of mining latent factors of drugs, and uses a novel balanced contrastive loss to optimize the parameters. \par

As shown in Equation (1) and (2), given drug $i$ and disease $j$, the BMF model utilizes the treated behavior of drug $i$,$R_{i*} \in \mathbb{R}^{n \times 1} $ and the treated behavior of the disease, $R_{*j} \in \mathbb{R}^{n \times 1} $ as the respective input instead of using one-hot coding, where $R \in \mathbb{R}^{m \times n} $ represents the drug-disease association matrix, $m$ denotes the number of drugs, $n$ denotes the number of diseases, $R_{i*}$ is the $i$th row of $R$, and $R_{*j}$ is the $j$th column of $R$. Subsequently, $R_{i*}$ and $R_{*j}$ are fed into a single-layer neural network for extracting the latent factor of drug $i$ and disease $j$. This operation is essentially embedding the behavioral information of drugs and diseases into their latent factors, which is used to enrich the expressiveness of the latent factors. \par

\begin{equation}
d^i = f(W_1^T R_{i*} + b_1)
\end{equation}

\begin{equation}
s^j = f(V_1^T R_{*j} + b_2)
\end{equation}

Where $W_1^T \in \mathbb{R}^{h \times n}$ and $V_1^T \in \mathbb{R}^{h \times m}$ are weight parameters, $h$ is the dimension of the latent factor vector. $b_1 \in \mathbb{R}^{h \times 1}$ and $b_2 \in \mathbb{R}^{h \times 1}$ are bias parameters. $f$ is the activation function, which can be ReLU or Sigmod function. $d^i \in \mathbb{R}^{h \times 1}$ and $s^j \in \mathbb{R}^{h \times 1}$ are the latent factor of drug $i$ and disease $j$, respectively. \par

Next, we do not simply calculate the similarity between the latent factors of the drug $i$ and the disease $j$ to obtain their compatibility. Rather, the representation of the target drug $i$ embeds the behavioral information of the target disease $j$. We treat all drugs that have therapeutic association with the target disease j as the set of neighbors of the target drug i,$N_i$. The set $N_i$ contains the latent factor of k diseases, and the value of k can be set manually. As shown in Equation (3), we first aggregate the all latent factors in the $N_i$ using average pooling. Where $d^t$ is the latent factor of the neighbor drug $t$. $o^i$ is the vector containing behavioral information after pooling (target disease $j$). This approach is straightforward, but has been proven effective in multiple scenarios. \par

\begin{equation}
o^i = \frac{1}{k} \sum_{t=1}^{k} d^t, \quad  d^t \in N_i
\end{equation}

We then combine the behavioral information vector $o^i$ and the latent factor of the target drug $i$ by Equation (4). \par

\begin{equation}
h^i = g d^i + (1-g) o^i
\end{equation}

Where $g$ is a manually defined weight parameter, and its value is between $0-1$. In this way, the BMF model successfully embeds multiple behavioral information (target drug and target disease) into the representation of the target drug $i$ ($h^i$). Finally, we use Equation (5) to predict the probability that drug $i$ can treat disease $j$. \par

\begin{equation}
\hat{R}_{ij} = g (W_2^T (h^i \odot s^j) + b_3)
\end{equation}

Where $W_2^T \in \mathbb{R}^{1 \times h}$ is the weight parameter, $\odot$ represents the Hadamard product, $\hat{R}_{ij}$ represents the probability of drug $i$ treating disease $j$. The parameters included in the above BMF model are optimized by the balanced contrastive loss described in the next section. \par

\subsection{Balanced Contrastive Loss}

The cross-entropy loss function is the dominant loss function in the field of computational drug repositioning. But it suffers from category imbalance problem, which leads to poor optimization results. In this work, we propose balance contrastive loss for overcoming the category imbalance problem. Given the validated drug-disease association pairs $(i,j)$, and the set of diseases $U$ that do not have a therapeutic relationship with the drug $i$ sampled by the negative sampling technique. The mathematical formula of BCL is shown in Equation (6). \par

\begin{equation}
\begin{aligned}
BCL & = -\alpha (1-\hat{R}_{ij})^\gamma \log \hat{R}_{ij} \\
& + \sum_{ u \in U} (1-\alpha)\hat{R}_{iu}^\gamma  \max(0, -\log(1-\hat{R}_{iu}+c) )
\end{aligned}
\end{equation}

Where $\hat{R}_{ij}$ denotes the predictive value of the BCF model, and $\alpha$ is used to balance the proportion of weights of positive and negative samples in the loss function. $\gamma$ is the adjustment coefficient, which is used to reduce the loss weights of the samples with high confidence. $c$ is the margin used to filter uninformative negative samples, and the value is usually set between 0 and 1. Specifically, BCL is used to optimize the parameters in BCL by maximizing the similarity between the target drug $i$ and positive disease $j$, and minimizing the similarity between the target drug $i$ and some negative diseases. \par

Next, we specifically  discuss why BCL can solve the category imbalance problem. Firstly, for the validated drug-disease association $(i,j)$, previous work uses negative sampling techniques to mine a large number of negative samples. As the number of samples increases, many redundant samples are usually generated. These samples do not contain valuable information for model training and typically have low predictive values. However, due to the advantage in the number of these samples, the accumulated gradients will affect the optimization direction of the model, reducing its generalization performance and convergence. In order to solve this problem, BCL uses the minimum margin, c, to filter the gradient of useless samples in the loss function. When the predicted value of the sample is lower than the minimum margin c, its loss will be set to 0. This makes these useless samples do not interfere with the optimization direction of the model. \par

Secondly, the cross-entropy loss function treats high-confidence samples and low-confidence samples equally. The imbalance between positive and negative samples results in a larger number of high-confidence samples, which occupies most of the loss function, and the direction of the gradient tends to further improve the confidence of high-confidence samples. However, these high-confidence samples improve the generalization performance of the model very little, and the model should focus on improving the confidence of low-confidence samples. Specifically, BCL reduces the loss weight of high-confidence samples through the parameter $\gamma$, so that the model focuses on low-confidence samples. It is worth noting that when the value of $\gamma$ is 1, BCL is equivalent to other loss functions, that is, to treat high-confidence samples and low-confidence samples equally. However, when the value of $\gamma$ is 2, if $\hat{r}_{iu})=0.01$, $0.01^2$ is approximately equal to 0.0001, while the loss weight of the traditional loss function is 1, and the loss weight is reduced by nearly 1000 times. In this way, the weight proportion of high-confidence samples in the loss function is reduced, so that the model can focus on improving low-confidence samples, thereby improving the generalization ability of the model. \par

Thirdly, the traditional loss function weights all samples as 1, but this strategy is inappropriate in category-imbalanced datasets. Because the number of drug-disease associations without treatment relationship is too large, it will dominate the direction of the optimization gradient, so that the model eventually tends to classify the samples to be predicted as drug-disease associations without treatment relationship. In the computational drug repositioning problem, we should pay more attention to the information brought by the validated drug-disease associations. Therefore, BCL adjusts the weight of negative samples in the loss function by the parameter $\alpha$, so that the model can learn more essential laws behind the drug-disease association from unbalanced data. \par

Figure 2 shows the change in latent factor (drug and disease) before and after training. It can be seen that BCL maximizes the similarity between drugs with therapeutic associations and diseases. And the gradient of the samples under the minimum margin will be set to 0, which will not interfere with the convergence direction of the model. Algorithm 1 is the pseudo-code of the BMF model. \par

\begin{algorithm}
	\caption{BMF.} 
	\textbf{Input-1:} Drug-Disease association matrix, $R$\\
	\textbf{Output:} Probability of treatment of disease $j$ by drug $i$, $ \hat{R}_{ij}$
	
	\begin{algorithmic}[1]
		
		\For {$(i,j)\in$ sampled drug-disease associations}

		\State \quad Using behavioral information to mine the latent factor of drug $i$.
		\State \quad \quad $	d^i \leftarrow f(W_1^T R_{i*} + b_1)$
		
		\State \quad Using behavioral information to mine the latent factor of disease $j$.
		\State \quad \quad $	s^j \leftarrow f(V_1^T R_{*j} + b_2)$
		
		\State \quad Aggregating the latent factors of neighboring drugs.
		\State \quad \quad $	o^i \leftarrow \frac{1}{k} \sum_{t=1}^{k} d^t, \quad  d^t \in N_i$
		
		\State \quad Combining the latent factor.
		\State \quad \quad $	h^i = g d^i + (1-g) o^i$
		
		\State \quad Calculating the predicted value.
		\State \quad \quad $	\hat{R}_{ij} = g (W_2^T (h^i \odot s^j) + b_3)$
		
		\State \quad Using BCL to optimize parameters in BMF model.
		\State \quad \quad $	BCL  = -\alpha (1-\hat{R}_{ij})^\gamma \log \hat{R}_{ij}$
		\State \quad \quad $	+ \sum_{ u \in U} (1-\alpha)\hat{R}_{iu}^\gamma  \max(0, -\log(1-\hat{R}_{iu}+c) )$

		\EndFor
		\State \textbf{Return $\hat{R}_{ij}$}

	\end{algorithmic} 
\end{algorithm}

\begin{table*}[h]
	\centering
	\caption{The experimental results of BMF model and BMF-OH model on three datasets}
	\setlength{\tabcolsep}{4mm}
	\begin{tabular}{@{}c|ccc|ccc|ccc@{}}
		\toprule
		Dataset & \multicolumn{3}{c|}{Gottlieb} & \multicolumn{3}{c|}{Cdataset} & \multicolumn{3}{c}{DNdataset} \\ \midrule
		& AUPR    & F1-Score   & HR     & AUPR    & F1-Score  & HR      & AUPR    & F1-Score  & HR      \\ \midrule
		BMF-OH  & 0.017   & 0.051      & 40\%   & 0.019   & 0.044     & 46.4\%  & 0.01    & 0.03      & 21.5\%  \\
		BMF     & 0.166   & 0.247      & 60\%   & 0.22    & 0.281     & 64.5\%  & 0.057   & 0.115     & 33.9\%  \\ \bottomrule
	\end{tabular}
\end{table*}

\begin{table*}[h]
	\centering
	\caption{The experimental results of BMF model and BMF-BCE model on three datasets}
	\setlength{\tabcolsep}{4mm}
	\begin{tabular}{@{}c|ccc|ccc|ccc@{}}
		\toprule
		Dataset & \multicolumn{3}{c|}{Gottlieb} & \multicolumn{3}{c|}{Cdataset} & \multicolumn{3}{c}{DNdataset} \\ \midrule
		& AUPR    & F1-Score  & HR      & AUPR    & F1-Score  & HR      & AUPR    & F1-Score  & HR      \\ \midrule
		BMF-BCE & 0.151   & 0.234     & 58.9\%  & 0.214   & 0.275     & 63.2\%  & 0.052   & 0.104     & 33\%    \\
		BMF     & 0.166   & 0.247     & 60\%    & 0.22    & 0.281     & 64.5\%  & 0.057   & 0.115     & 33.9\%  \\ \bottomrule
	\end{tabular}
\end{table*}

\section{Experiments and Discussion}

In this section, we discuss the following research questions for the BMF model.

\begin{enumerate}
	
	\item[\textbf{RQ1}] Does the operation of embedding behavioral information into the BMF model to enhance the drug representation contribute to the model's performance?
	
	\item[\textbf{RQ2}] Does BCL improve the performance of the model in category imbalance scenarios compared to the cross-entropy loss function?
	
	\item[\textbf{RQ3}]  Does the BMF model outperform the current SOTA model?
\end{enumerate}

\begin{figure}[t]
	\centering

	\subfigure[Fdataset (P-R Curves)]{
		\includegraphics[width=6cm,height=5cm]{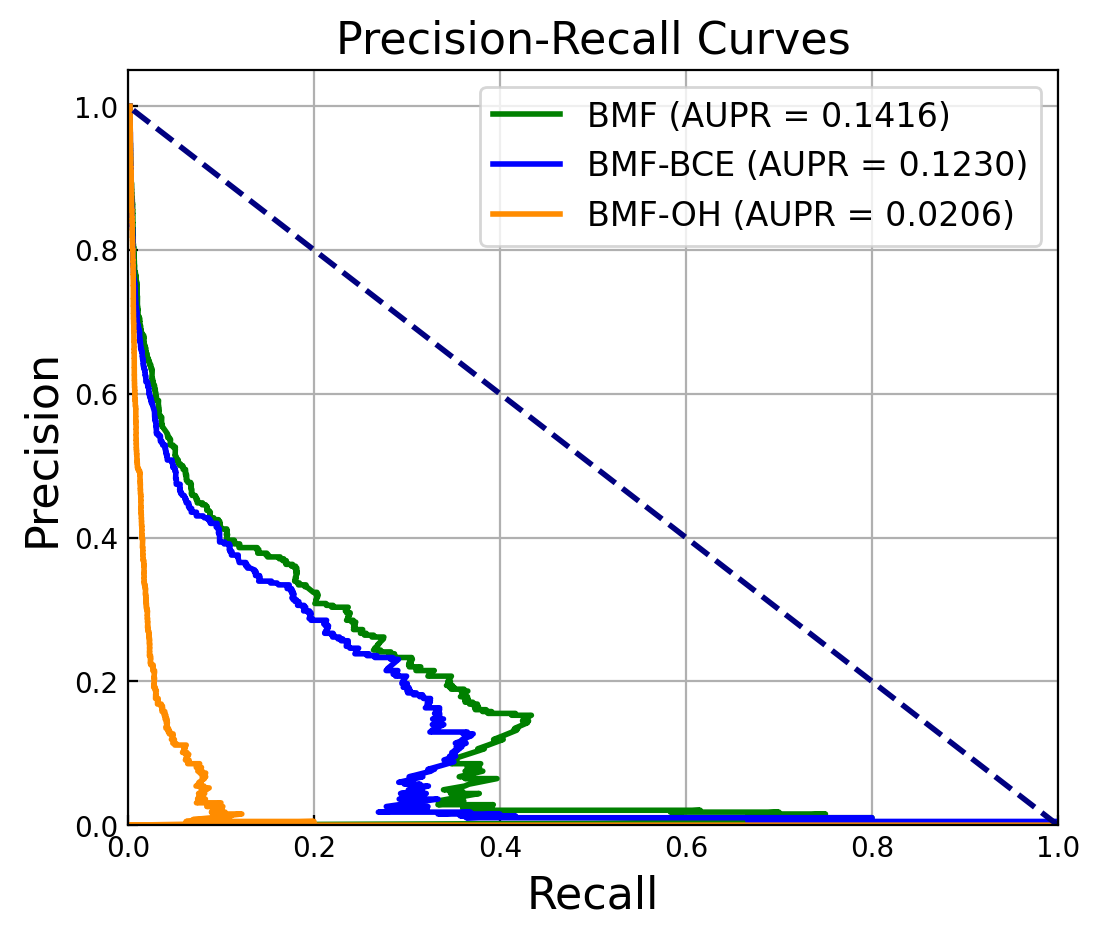}
	}

	\subfigure[Cdataset (P-R Curves)]{
		\includegraphics[width=6cm,height=5cm]{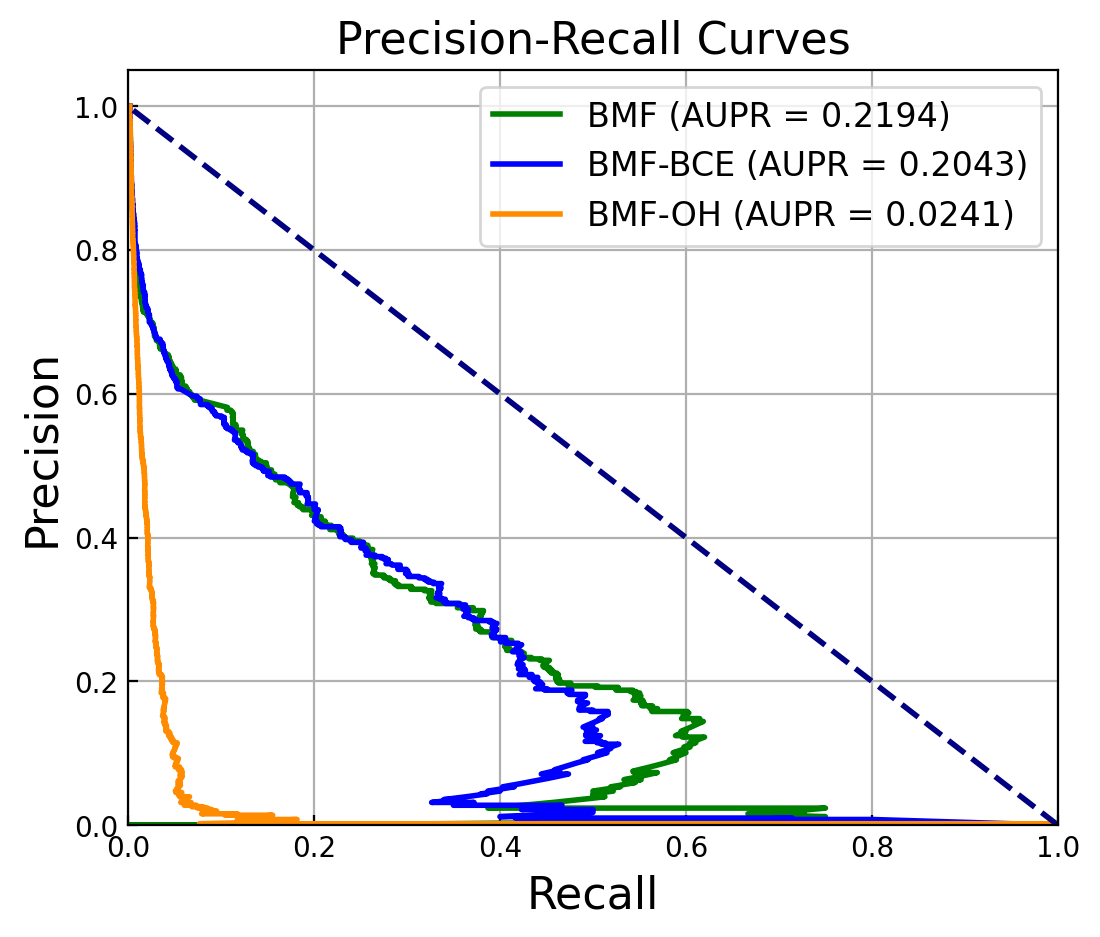}
	}

	\subfigure[DNdataset (P-R Curves)]{
		\includegraphics[width=6cm,height=5cm]{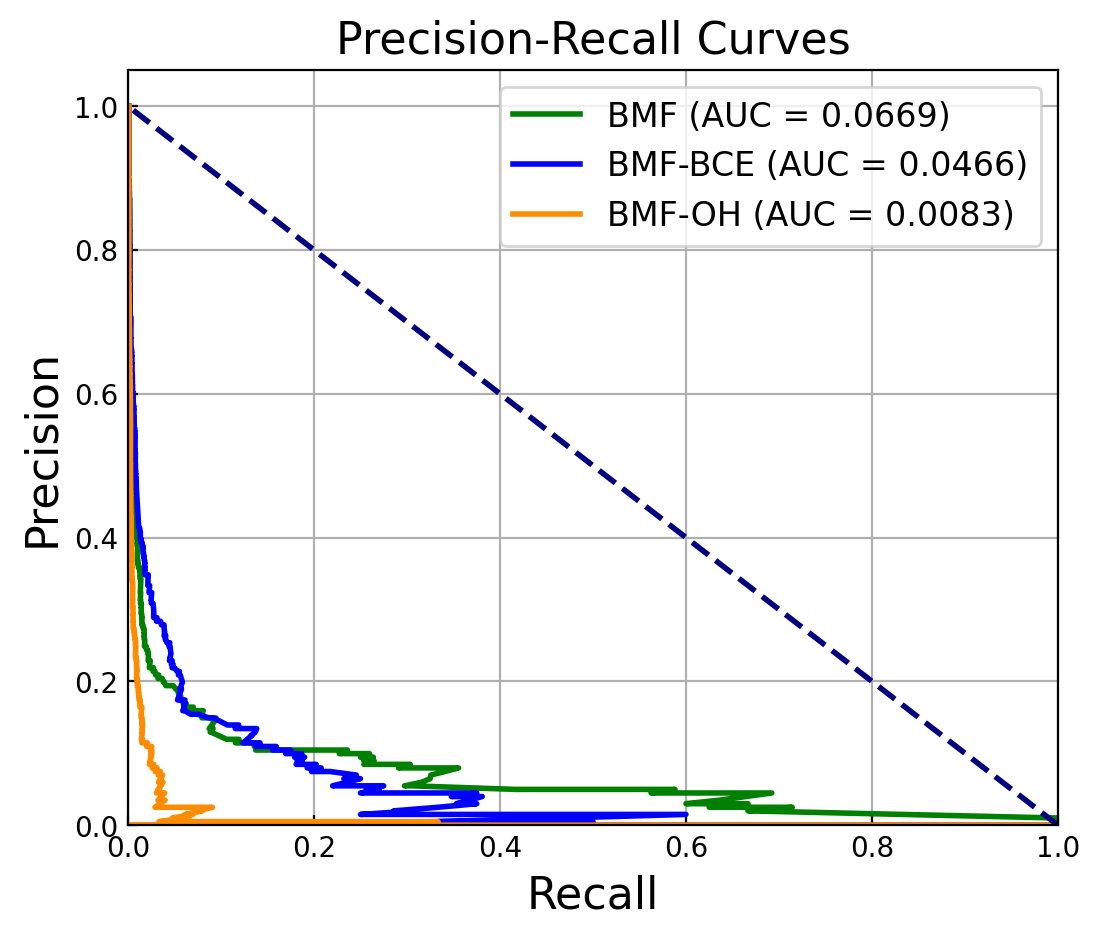}
	}

	\caption{ The P-R Curves of BMF, BMF-BCE and BMF-OH model.}
\end{figure}\par

\subsection{The Setting of Evaluation Metrics}

In this study, three publicly available benchmark datasets were used to evaluate the BMF model. For each benchmark dataset, we employ 5-fold cross-validation to evaluate the BMF model as well as the benchmark method. And all models are trained and predicted on already cut data, which ensures the consistency of training and testing data and thus enables a fairer comparison of model performance. We use three popular evaluation metrics to numerically represent the performance of all models. These three evaluation metrics are AUPR (Area Under Precision-Recall Curve), F1-score and Hit Ratio (HR).\par

\begin{table*}[t]
	\centering
	\caption{The experimental results of the BMF model with all the comparison algorithms}
	\setlength{\tabcolsep}{4mm}
	\begin{tabular}{@{}c|ccc|ccc|ccc@{}}
		\toprule
		Dataset & \multicolumn{3}{c|}{Gottlieb} & \multicolumn{3}{c|}{Cdataset} & \multicolumn{3}{c}{DNdataset} \\ \midrule
		& AUPR    & F1-Score  & HR      & AUPR    & F1-Score  & HR      & AUPR    & F1-Score  & HR      \\ \midrule
		MF      & 0.033   & 0.01      & 22.1\%  & 0.021   & 0.01      & 22.4\%  & 0.005   & 0.003     & 10.5\%  \\
		NCF     & 0.022   & 0.051     & 40.8\%  & 0.023   & 0.046     & 45\%    & 0.009   & 0.037     & 15.1\%  \\
		MLP     & 0.021   & 0.05      & 40\%    & 0.027   & 0.049     & 41.4\%  & 0.007   & 0.038     & 14\%    \\
		MetricF & 0.039   & 0.073     & 43.7\%  & 0.092   & 0.091     & 50.7\%  & 0.006   & 0.028     & 21.9\%  \\
		ADAE     & 0.021   & 0.63      & 37.6\%  & 0.041   & 0.076     & 44.6\%  & 0.039   & 0.054     & 25.4\%  \\
		BMF-OH  & 0.017   & 0.051      & 40\%   & 0.019   & 0.044     & 46.4\%  & 0.01    & 0.03      & 21.5\%  \\
		BMF-BCE & 0.151   & 0.234     & 58.9\%  & 0.214   & 0.275     & 63.2\%  & 0.052   & 0.104     & 33\%    \\
		BMF     & 0.166   & 0.247     & 60\%    & 0.23    & 0.281     & 64.5\%  & 0.057   & 0.115     & 33.9\%  \\ \bottomrule
	\end{tabular}
\end{table*}

\subsection{Hyperparameter Setup}

The hyperparameters in the BMF model include the length of the latent factor, $h$, which affects the expression ability of the latent factor of the drug or disease; The size of the neighborhood drug set $N_i$ in Equation (3), the larger the value, the greater the number of drugs contained in the set, the more comprehensive the information while also increasing the amount of computation; The parameter $g$ in Equation (4), which is used to regulate the weight of behavioral information in the drug representation. The parameters $\alpha$, $\gamma$ and $c$ in Equation (6), $\alpha$ is used to adjust the proportion of positive and negative samples in the loss function, $\gamma$ represents the penalty strength of the loss function for high-confidence samples, and $c$ represents the minimum filtering margin. The variation intervals of the above hyperparameters are [8,16,64,128,256,512], [1,5,10,20], [0.1,0.3,0.5,0.7,0.9], [0.1,0.3,0.5,0.7,0.9], [1,2,3] and [0.001,0.01,0.1,0.2], respectively. \par

\subsection{Effectiveness of the Behavioral Information (RQ1)}

In this study, we argue that using one-hot coding as input to mine the latent factor of drug reduces the expressive ability of the model. The one-hot coding only contains information about the location of the drug, which lacks modeling of behavioral information. The behavioral information has been shown to be an effective way to improve model accuracy in recommender systems. Therefore, we take the drug's therapeutic behavior as input to mine its latent factor, and additionally incorporate information from neighboring drugs. To verify whether the above behavioral information can outperform the traditional one-hot coding, we experimentally compare the BMF model with its variant version BMF-OH, where the input of the latter is one-hot coding. \par

Table \uppercase\expandafter{\romannumeral2} shows the experimental results of BMF and BMF0-OH on three datasets. The experimental results intuitively show that the BMF model outperforms the BMF-OH model on all three datasets. Specifically, in Fdataset, the AUPR value of BMF-OH is 0.017 and that of BMF is 0.166 with an improvement of 876\%; The F1-Score value of BMF-OH is 0.051 and that of BMF is 0.247 with an improvement of 384\%; The HR value of BMF-OH is 40\% and the HR value of BMF is 60\% with an improvement of 49.9\%. Furthermore, the average improvement on Cdataset and DNdataset is 402\% and 48.8\%, respectively. \par

By analyzing the experimental results of the BMF model and BMF-OH, we can conclude that modeling the behavior information of the drug can enhance the expression of its latent factor, thereby improving the prediction accuracy of the model. \par

\begin{figure*}[t]
	\centering
	
	\subfigure[Fdataset]{
		\includegraphics[width=5cm,height=4cm]{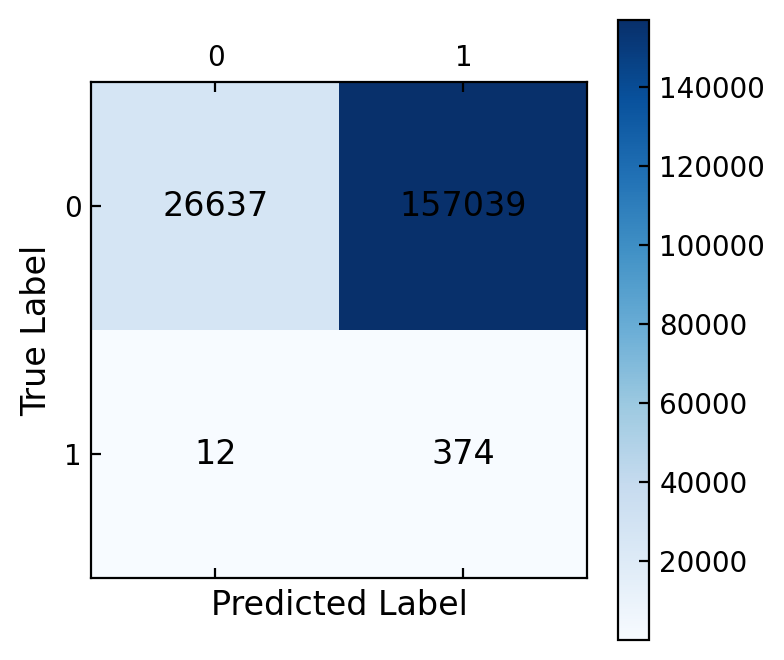}
	}
	\quad
	\subfigure[Cdataset]{
		\includegraphics[width=5cm,height=4cm]{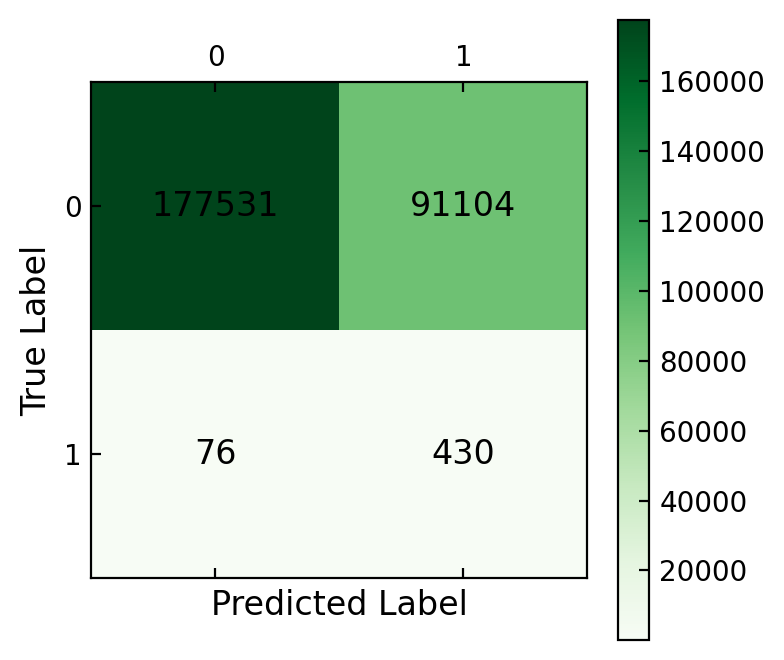}
	}
	\quad
	\subfigure[DNdataset]{
		\includegraphics[width=5cm,height=4cm]{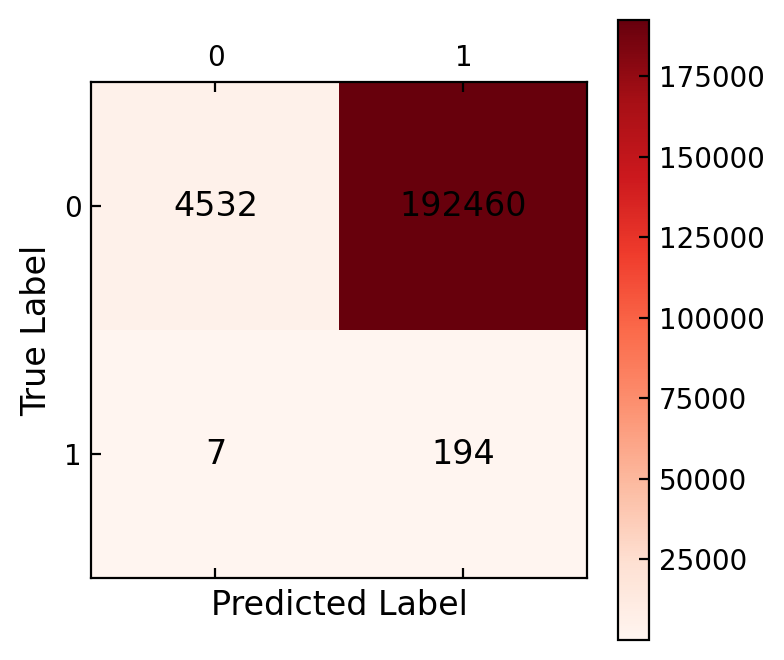}
	}
	
	\caption{ The confusion matrix of BMF model when threshold = 0.1}
\end{figure*}\par

\begin{figure*}[t]
	\centering
	
	\subfigure[Fdataset]{
		\includegraphics[width=5cm,height=4cm]{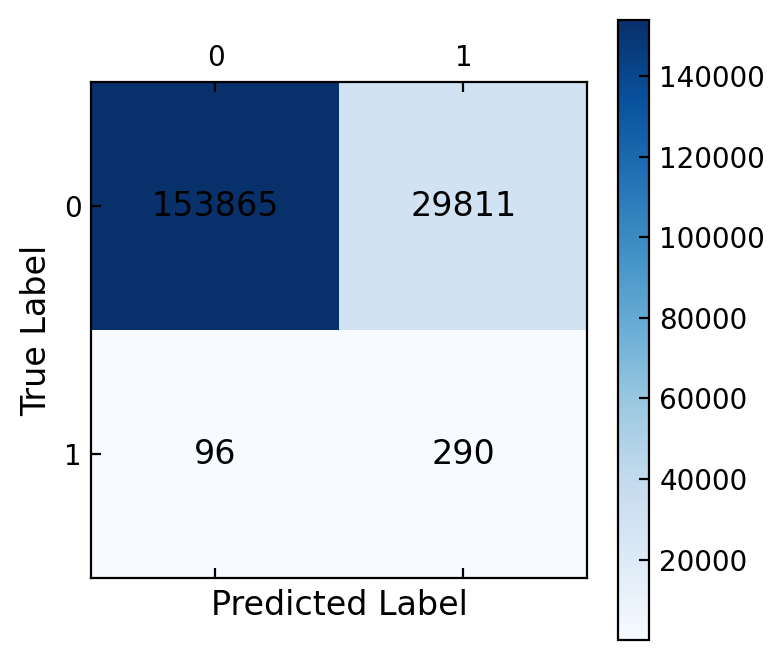}
	}
	\quad
	\subfigure[Cdataset]{
		\includegraphics[width=5cm,height=4cm]{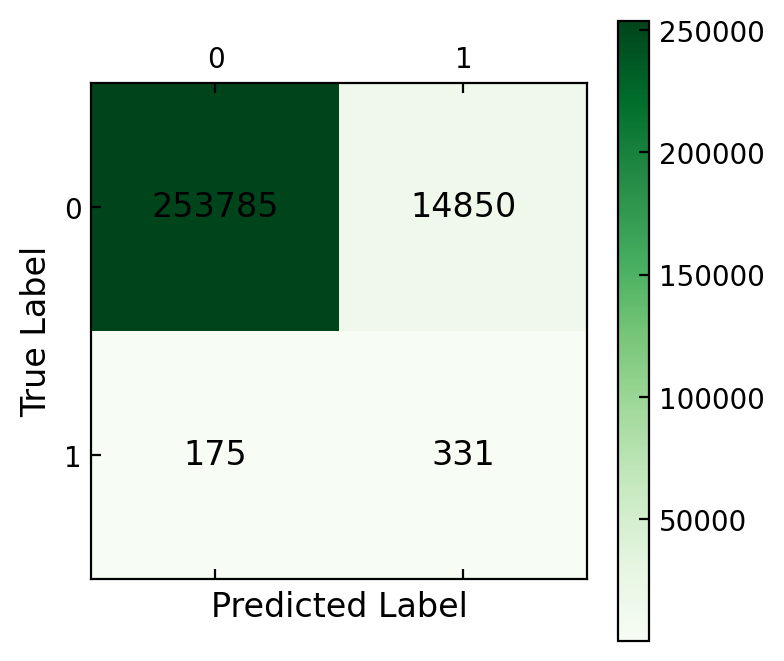}
	}
	\quad
	\subfigure[DNdataset]{
		\includegraphics[width=5cm,height=4cm]{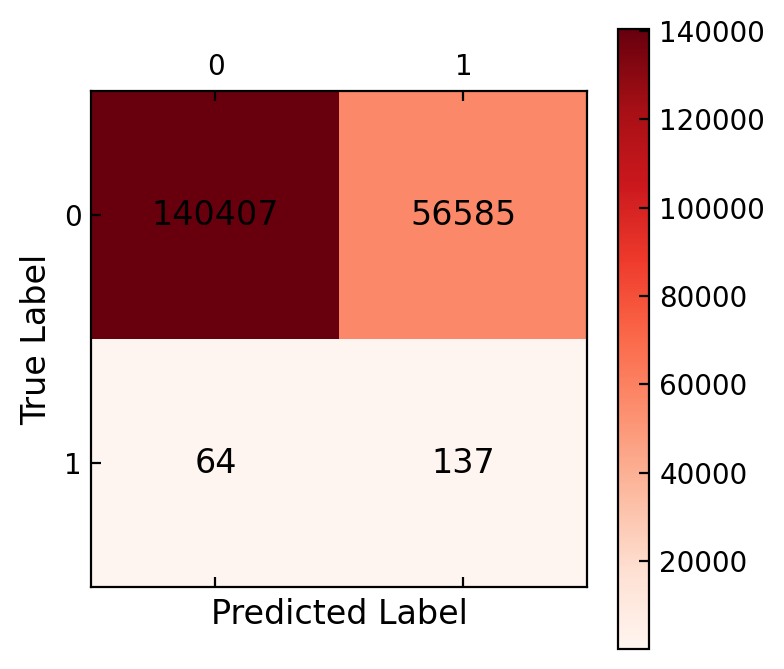}
	}
	
	\caption{ The confusion matrix of BMF model when threshold = 0.3}
\end{figure*}

\subsection{Effectiveness of the Balanced Contrastive Loss (RQ2)}

Previous loss functions are not effective in dealing with the category imbalance problem in computational drug repositioning, so we propose a balanced contrast loss (BCL) in this study. The advantage of BCL is that it places more emphasis on low confidence samples, uses minimum margin to filter uninformative samples, and dynamically adjusts the weights of positive and negative samples in the loss function. In order to investigate the effectiveness of BCL, we set up the following comparison experiment. This comparison experiment compares BMF with its variant version BMF-BCE, where the difference between the two is that the latter uses the binary cross-entropy loss function. \par

Table \uppercase\expandafter{\romannumeral3} shows the experimental results of BMF and BMF-BCE on the three datasets. The above experimental results intuitively show that the BMF model outperforms the BMF-BCE model on all three data sets. Specifically, the improvement in Fdataset, Cdataste, and DNdataset are 5.7\%, 3.9\%, and 7.6\%, respectively. \par

By analyzing the above experimental results, we can easily conclude that the BCL can better solve the problem of category imbalance in computational drug repositioning than the binary cross-entropy loss function. \par

\subsection{Baseline Comparison (RQ3)}

The following is a comprehensive performance comparison of BMF models with currently popular models on three publicly available datasets. To highlight the superiority of the BMF models, we use the same dataset partitioning method and evaluation metrics for all models. The description of the benchmark model used for the experiments is as follows. \par

\begin{itemize}

	\item \emph{MF} \cite{rw21}: The matrix factorization model (MF) uses the latent factor of the drug and the disease to perform the inner product operation to calculate the treatment probability between the drug and the disease.
	\item \emph{NCF} \cite{ncf}: The neural collaborative filtering model (NCF) takes the Hadamard product of the latent factors of the drug and the disease, and then inputs it into the neural network to calculate the treatment probability between the drug and the disease.
	\item \emph{MLP} \cite{mlp}: The multilayer perceptrons (MLP) are composed of multiple neural networks, which can effectively capture the nonlinear association between drugs and diseases.
	
	\item \emph{MetricF} \cite{metricf}: The metric factorization model (MetricF) treats the latent factor of drug and disease as a point in the Euclidean space, and subsequently uses Euclidean distance to represent the relationship between the drug and the disease.
	\item \emph{ADAE} \cite{adae}: The additional denoising autoencoder (ADAE) mines the latent factors of drugs and diseases based on the idea that good quality latent factors can restore the original input, and additionally considers the similarity information of drugs and diseases. The probability of a therapeutic relationship between the drug and the disease is then calculated using an inner product operation.

\end{itemize}

Table \uppercase\expandafter{\romannumeral4} shows the experimental results of the BMF model with the benchmark models on the three datasets. We can obtain several obvious conclusions as follows. Firstly, we find that the BMF model achieves the best prediction results on Fdataset, Cdataset, and DNdataset. Specifically, the BMF model achieves values of 0.166, 0.247, and 60\% for the AUPR, F1-Score, and HR metrics on Fdataset, respectively. The values of the BMF model on Cdataset for AUPR, F1-Score, and HR metrics are 0.23, 0.281, and 64.5\%, respectively. The values of the BMF model on DNdataset for AUPR, F1-Score, and HR metrics were 0.057, 0.115, and 33.9\%, respectively. \par

Furthermore, MF, as the basis for latent factor-based models, achieves the worst predictive performance. And NCF obtains better results than the MF model because it considers the weights of different dimensions in the latent factor vector. And the experimental results of the MetricF model are also better than MF, which also verifies that the expression of point and distance is better than the inner product and vector, making up for the defect that it cannot satisfy the triangle inequality. \par

In addition, Figure 3 shows the Precision-Recall (P-R) Curves of BMF, BMF-BCE, and BMF-OH models. We find that under the P-R curve, the BMF model outperforms the BMF-BCE model. The P-R curve is a mainstream evaluation metric for binary classification problems, which can fairly compare the performance of models on category imbalanced dataset. Therefore, the performance of the BMF model and the BMF-BCE model under the P-R curve verifies the effectiveness of the balanced contrastive loss proposed in this work. Another point, under the P-R curve, the performance of the BMF model is better than the BMF-OH model. This illustrates the validity of the behavioral information proposed in this work. \par

Figures 4 and 5 show the confusion matrix of BMF at different thresholds. On the one hand, we find that the BMF model has a high probability of correctly classifying positive samples when the threshold is 0.1. However, it also has a higher probability of predicting the true negative samples as positive samples. On the other hand, when the threshold is 0.3, the BMF model has a lower probability of correctly classifying positive samples, but a higher probability of correctly classifying negative samples. Therefore, we need to dynamically adjust the threshold to make the model's prediction results meet the needs of real scenarios. \par

After the above discussion, we can find that the BMF model is in the leading position in terms of predictive performance, indicating that the use of BCL can better cope with the problem of category imbalance in computational drug repositioning, and that modeling the behavioral information of drugs can likewise yield a better drug representation. These findings are a direct indication of the effectiveness of these two improvement points in this work. \par

\section{Conclusion}

In this work, we propose a balanced matrix factorization with embedded behavior information (BMF) for computational drug repositioning to address the above-mentioned shortcomings. The BMF model uses behavioral information to enhance drug representation, and proposes a BCL loss function for improving the optimization direction of the model under the category imbalance problem, allowing better convergence of the model parameters. The effectiveness of these innovations is verified on ablation experiments on three real datasets. And the comparison with popular benchmark models verifies the superiority of BMF model. However, BMF, as a matrix factorization-based model, lacks the consideration of neighborhood structure information, thus limiting its generalization ability. Therefore, in future work, we will explore how to embed the neighborhood structure information into the BMF model.



\begin{thebibliography}{00}
	
	\bibitem{1} M. Dickson and J. P. Gagnon, “Key factors in the rising cost of new drug discovery and development,” \emph{Nature Reviews Drug discovery}, vol. 3, no. 5, pp. 417–429, 2004.
	
	\bibitem{2} N. A. Tamimi and P. Ellis, “Drug development: from concept to marketing!” \emph{Nephron Clinical Practice}, vol. 113, no. 3, pp. c125–c131, 2009.
	
	\bibitem{3} T. T. Ashburn and K. B. Thor, “Drug repositioning: identifying and developing new uses for existing drugs,” \emph{Nature Reviews Drug Discovery}, vol. 3, no. 8, pp. 673–683, 2004.
	
	
	\bibitem{4} N. Nosengo, “Can you teach old drugs new tricks?” \emph{Nature News},
	vol. 534, no. 7607, p. 314, 2016.
	
	\bibitem{5} J.-L. E. Pritchard, T. A. O’Mara, and D. M. Glubb, “Enhancing
	the promise of drug repositioning through genetics,” \emph{Frontiers in
		Pharmacology}, vol. 8, p. 896, 2017.
	
	\bibitem{6} M. Lotfi Shahreza, N. Ghadiri, S. R. Mousavi, J. Varshosaz, and
	J. R. Green, “A review of network-based approaches to drug
	repositioning,” \emph{Briefings in Bioinformatics}, vol. 19, no. 5, pp. 878–
	892, 2018.
	
	\bibitem{7} B. Padhy, Y. Gupta et al., “Drug repositioning: re-investigating existing drugs for new therapeutic indications,” \emph{Journal of Postgraduate Medicine}, vol. 57, no. 2, p. 153, 2011.
	
	\bibitem{8}  S. S. Sadeghi and M. R. Keyvanpour, “An analytical review
	of computational computational drug repositioning,” \emph{IEEE/ACM Transactions on
		Computational Biology and Bioinformatics}, vol. 18, no. 2, pp. 472–488,
	2019.
	
	\bibitem{9} J. K. Yella, S. Yaddanapudi, Y. Wang, and A. G. Jegga, “Changing trends in computational drug repositioning,” \emph{Pharmaceuticals}, vol. 11, no. 2, p. 57, 2018.
	
	\bibitem{10} H. Luo, M. Li, M. Yang, F.-X. Wu, Y. Li, and J. Wang, “Biomedical data and computational models for drug repositioning: a comprehensive review,” \emph{Briefings in Bioinformatics}, vol. 22, no. 2, pp. 1604–1619, 2021.
	
	\bibitem{11} A. Gottlieb, G. Y. Stein, E. Ruppin, and R. Sharan, “Predict: a
	method for inferring novel drug indications with application to
	personalized medicine,” \emph{Molecular Systems Biology}, vol. 7, no. 1, p.
	496, 2011.
	
	\bibitem{12} J. Yang, D. Zhang, L. Liu, G. Li, Y. Cai, Y. Zhang, H. Jin, and X. Chen, “Computational drug repositioning based on the relationships between substructure–indication,” \emph{Briefings in Bioinformatics}, vol. 22, no. 4, p. bbaa348, 2021.
	
	\bibitem{13} H. Luo, M. Li, S. Wang, Q. Liu, Y. Li, and J. Wang, “Computational
	drug repositioning using low-rank matrix approximation and
	randomized algorithms,” \emph{Bioinformatics}, vol. 34, no. 11, pp. 1904–
	1912, 2018.
	
	\bibitem{14} M. Yang, G. Wu, Q. Zhao, Y. Li, and J. Wang, “Computational
	drug repositioning based on multi-similarities bilinear matrix
	factorization,” \emph{Briefings in Bioinformatics}, vol. 22, no. 4, p. bbaa267,
	2021.
	
	\bibitem{15} X. Yang, Y. Liu, J. He et al., “Additional neural matrix factorization
	model for computational drug repositioning,” \emph{BMC bioinformatics},
	vol. 20, no. 1, pp. 1–11, 2019.
	
	\bibitem{16} J. He, X. Yang, Z. Gong et al., “Hybrid attentional memory network
	for computational drug repositioning,” \emph{BMC bioinformatics}, vol. 21,
	no. 1, pp. 1–17, 2020.
	
	
	\bibitem{n1} W. Wang, S. Yang, X. Zhang, and J. Li, “Drug repositioning by
	integrating target information through a heterogeneous network
	model,” \emph{Bioinformatics}, vol. 30, no. 20, pp. 2923–2930, 2014.
	
	
	\bibitem{n2} X. Zeng, S. Zhu, X. Liu, Y. Zhou, R. Nussinov, and F. Cheng,
	“deepdr: a network-based deep learning approach to in silico drug
	repositioning,” \emph{Bioinformatics}, vol. 35, no. 24, pp. 5191–5198, 2019.
	
	\bibitem{n3} Y. Meng, C. Lu, M. Jin, J. Xu, X. Zeng, and J. Yang, “A weighted
	bilinear neural collaborative filtering approach for drug repositioning,” \emph{Briefings in Bioinformatics}, 2022.
	
	\bibitem{rw11} H. Luo, J. Wang, M. Li, J. Luo, X. Peng, F.-X. Wu, and Y. Pan, “Drug repositioning based on comprehensive similarity measures and bi-random walk algorithm,” \emph{Bioinformatics}, vol. 32, no. 17, pp.
	2664–2671, 2016.
	
	\bibitem{rw12} H. Chen, F. Cheng, and J. Li, “idrug: Integration of drug repositioning and drug-target prediction via cross-network embedding,”
	\emph{PLoS Computational Biology}, vol. 16, no. 7, p. e1008040, 2020.
	
	
	\bibitem{rw13} L. Cai, C. Lu, J. Xu, Y. Meng, P. Wang, X. Fu, X. Zeng, and Y. Su,
	“Drug repositioning based on the heterogeneous information fusion
	graph convolutional network,” \emph{Briefings in Bioinformatics}, vol. 22,
	no. 6, p. bbab319, 2021.
	
	\bibitem{rw14} Y. Ge, T. Tian, S. Huang, F. Wan, J. Li, S. Li, X. Wang, H. Yang,
	L. Hong, N. Wu et al., “An integrative drug repositioning framework discovered a potential therapeutic agent targeting covid-19,” \emph{Signal Transduction and Targeted Therapy}, vol. 6, no. 1, pp. 1–16, 2021.
	
	\bibitem{m1} W. Zhang, H. Xu, X. Li, Q. Gao, and L. Wang, “Drimc: an improved
	drug repositioning approach using bayesian inductive matrix
	completion,” \emph{Bioinformatics}, vol. 36, no. 9, pp. 2839–2847, 2020.
	
	\bibitem{m2} Y. Yan, M. Yang, H. Zhao, G. Duan, X. Peng, and J. Wang,
	“Drug repositioning based on multi-view learning with matrix
	completion,” \emph{Briefings in Bioinformatics}, vol. 23, no. 3, p. bbac054,
	2022.
	
	\bibitem{m3} X. Yang, G. Yang, and J. Chu, “The neural metric factorization
	for computational drug repositioning,” \emph{IEEE/ACM Transactions on
		Computational Biology and Bioinformatics}, 2022.
	
	\bibitem{rw21} Y. Koren, “Factorization meets the neighborhood: a multifaceted
	collaborative filtering model,” \emph{in Proceedings of the 14th ACM
		SIGKDD International Conference on Knowledge Discovery and Data
		Mining}, 2008, pp. 426–434.
	
	
	\bibitem{rw22} P. Chen, T. Bao, X. Yu, and Z. Liu, “A drug repositioning algorithm based on a deep autoencoder and adaptive fusion,” \emph{BMC Bioinformatics}, vol. 22, no. 1, pp. 1–18, 2021.
	
	
	\bibitem{rw23} H. Luo, J. Wang, C. Yan, M. Li, F.-X. Wu, and Y. Pan, “A novel drug
	repositioning approach based on collaborative metric learning,”
	\emph{IEEE/ACM Transactions on Computational Biology and Bioinformatics},
	vol. 18, no. 2, pp. 463–471, 2019.
	
	
	\bibitem{simplex} K. Mao, J. Zhu, J. Wang, Q. Dai, Z. Dong, X. Xiao, and X. He,
	“Simplex: A simple and strong baseline for collaborative filtering,”
	\emph{in Proceedings of the 30th ACM International Conference on Information
		\& Knowledge Management}, 2021, pp. 1243–1252.
	
	\bibitem{focaloss} T.-Y. Lin, P. Goyal, R. Girshick, K. He, and P. Dollár, “Focal loss
	for dense object detection,” \emph{in Proceedings of the IEEE International
		Conference on Computer Vision}, 2017, pp. 2980–2988.
	
	
	\bibitem{cl} F. Wang and H. Liu, “Understanding the behaviour of contrastive
	loss,” \emph{in Proceedings of the IEEE/CVF Conference on Computer Vision
		and Pattern Recognition}, 2021, pp. 2495–2504.
	
	\bibitem{drugbank} C. Knox, V. Law, T. Jewison, P. Liu, S. Ly, A. Frolkis, A. Pon,
	K. Banco, C. Mak, V. Neveu et al., “Drugbank 3.0: a comprehensive
	resource for ‘omics’ research on drugs,” \emph{Nucleic Acids Research}, vol. 39,
	no. suppl-1, pp. D1035–D1041, 2010.
	
	\bibitem{omim} A. Hamosh, A. F. Scott, J. S. Amberger, C. A. Bocchini, and
	V. A. McKusick, “Online mendelian inheritance in man (omim), a
	knowledgebase of human genes and genetic disorders,” \emph{Nucleic
		Acids Research}, vol. 33, no. suppl-1, pp. D514–D517, 2005.
	
	
	\bibitem{cdk} C. Steinbeck, Y. Han, S. Kuhn, O. Horlacher, E. Luttmann, and
	E. Willighagen, “The chemistry development kit (cdk): An opensource java library for chemo-and bioinformatics,” \emph{Journal of Chemical Information and Computer Sciences}, vol. 43, no. 2, pp. 493–500,
	2003.
	
	\bibitem{smile} D. Weininger, “Smiles, a chemical language and information system.
	1. introduction to methodology and encoding rules,” \emph{Journal of Chemical Information and Computer Sciences}, vol. 28, no. 1, pp. 31–36, 1988.
	
	
	\bibitem{ncf} X. He, L. Liao, H. Zhang, L. Nie, X. Hu, and T.-S. Chua, “Neural
	collaborative filtering,” \emph{in Proceedings of the 26th International Conference on World Wide Web}, 2017, pp. 173–182.
	
	
	\bibitem{mlp} Y. LeCun, Y. Bengio, and G. Hinton, “Deep learning,” \emph{Nature}, vol.
	521, no. 7553, pp. 436–444, 2015.
	
	\bibitem{metricf} S. Zhang, L. Yao, B. Wu, X. Xu, X. Zhang, and L. Zhu, “Unraveling
	metric vector spaces with factorization for recommendation,” \emph{IEEE
		Transactions on Industrial Informatics}, vol. 16, no. 2, pp. 732–742,
	2019.
	
	\bibitem{adae} X. Dong, L. Yu, Z. Wu, Y. Sun, L. Yuan, and F. Zhang, “A hybrid
	collaborative filtering model with deep structure for recommender
	systems,” \emph{in Proceedings of the AAAI Conference on Artificial Intelligence}, vol. 31, no. 1, 2017.
	
	
	
\end{thebibliography}
\end{document}